%% file: acl_latex.tex
\pdfoutput=1
\documentclass[11pt]{article}

\usepackage[]{acl}

\usepackage{times}
\usepackage{latexsym}

\usepackage[T1]{fontenc}
\usepackage[utf8]{inputenc}

\usepackage{microtype}
\usepackage{multirow}
\usepackage[T1]{fontenc}
\usepackage{subfigure}
\usepackage{graphicx}
\usepackage{amsmath}
\usepackage{amsfonts}
\usepackage{enumitem}
\usepackage{algorithm}
\usepackage{algorithmic}
\usepackage{booktabs}
\input{math_commands}

\usepackage{times}
\usepackage{latexsym}
\title{ViT-TTS: Visual Text-to-Speech with Scalable Diffusion Transformer}

\author{%
  Huadai Liu$^{1}$\thanks{~~Equal contributions}, Rongjie Huang$^{1}$\footnotemark[1], Xuan Lin$^{2}$\footnotemark[1], Wenqiang Xu$^{2}$, Maozong Zheng$^{2}$,  Hong Chen$^{2}$, \\ \textbf{Jinzheng He$^{1}$, Zhou Zhao$^{1}$\thanks{~~Corresponding author}} \\
  Zhejiang University$^1$, Ant Group$^2$ \\ \
  \texttt{\{liuhuadai,rongjiehuang,jinzhenghe,zhaozhou\}@zju.edu.cn} \\
    \texttt{\{daxuan.lx,yugong.xwq,zhengmaozong.zmz,wuyi.ch\}@antgroup.com}
  }

\begin{document}
\maketitle
\input{sections/abstract}
\input{sections/introduction}

\input{sections/related_work}
\input{sections/method}

\input{sections/experiment}

\input{sections/conclusion}
\bibliography{emnlp2023}
\clearpage
\appendix
\input{sections/appendix}

\end{document}

%% file: math_commands.tex
\usepackage{amsmath,amsfonts,bm}

\def\eqref#1{equation~\ref{#1}}
\def\1{\bm{1}}

\newcommand{\beps}{\boldsymbol\epsilon}

\def\rvepsilon{{\mathbf{\epsilon}}}

\def\rmI{{\mathbf{I}}}

\def\vzero{{\bm{0}}}

\def\vx{{\bm{x}}}

\def\vz{{\bm{z}}}

\def\mI{{\bm{I}}}

\DeclareMathAlphabet{\mathsfit}{\encodingdefault}{\sfdefault}{m}{sl}
\SetMathAlphabet{\mathsfit}{bold}{\encodingdefault}{\sfdefault}{bx}{n}

\def\gL{{\mathcal{L}}}

\def\gN{{\mathcal{N}}}

%% file: sections/abstract.tex
\begin{abstract}
Text-to-speech(TTS) has undergone remarkable improvements in performance, particularly with the advent of Denoising Diffusion Probabilistic Models (DDPMs). However, the perceived quality of audio depends not solely on its content, pitch, rhythm, and energy, but also on the physical environment.
In this work, we propose ViT-TTS, the first visual TTS model with scalable diffusion transformers. ViT-TTS complement the phoneme sequence with the visual information to generate high-perceived audio, opening up new avenues for practical applications of AR and VR to allow a more immersive and realistic audio experience. To mitigate the data scarcity in learning visual acoustic information, we 1) introduce a self-supervised learning framework to enhance both the visual-text encoder and denoiser decoder; 2) leverage the diffusion transformer scalable in terms of parameters and capacity to learn visual scene information. Experimental results demonstrate that ViT-TTS achieves new state-of-the-art results, outperforming cascaded systems and other baselines regardless of the visibility of the scene. With low-resource data (1h, 2h, 5h), ViT-TTS achieves comparative results with rich-resource baselines.~\footnote{Audio samples are available at \url{https://ViT-TTS.github.io/.}}~\footnote{Code is available at \url{https://github.com/liuhuadai/ViT-TTS/}}
% Specifically, we use a large volume of unlabeled data to pre-train the encoder and decoder to mitigate the scarcity of data via masked prediction loss. Additionally, we show that the WaveNet inductive bias is not crucial to the performance of DDPMs and they can readily be replaced with transformers. With the scaling prosperity of transformers, model capacity can be effortlessly expanded to better model room acoustic data and generate high-quality audio that matches the target environment. Experimental results demonstrate that ViT-TTS outperforms WaveNet backbone and cascaded model(composed of TTS and visual acoustic matching) in the Visual TTS task, achieving state-of-the-art performance and significant results in low-resource scenarios (1h/2h/5h).
% To address this, we propose the Visual TTS task, which involves converting given written text and target environmental images into audio that matches the target environment. Despite benefits, this task presents challenges: 1) the high cost of annotating high-quality target environmental images and audio leads to a scarcity of data for model training, and 2) modeling room acoustic information and integrating it with the audio properties to produce realistic target environmental audio is challenging.
\end{abstract}

%% file: sections/introduction.tex
% \begin{figure*}
%     \centering
%     \vspace{-4mm}
%     \includegraphics[scale=0.5]{figs/fig1.png}
%     \caption{Examples of panoramic image in SoundSpaces.}
%     \vspace{-4mm}
%     \label{fig:fig1}
% \end{figure*}
\section{Introduction}
% The sounds that we perceive are transmitted through various mediums, including solids, liquids, and gases, all of which have a significant impact on our auditory perception. 
Text-to-speech (TTS)~\citep{ren2019fastspeech,huang2022fastdiff,huanggenerspeech} aims to synthesize audios that is consistent with the reference samples in terms of semantic meaning, timbre, emotions, and melody, and has shown remarkable advancements with the advent of Denoising Diffusion Probabilistic Models (DDPMs). However, the perceived audio quality is not solely determined by these aspects, as it is also influenced by the surrounding physical environment.
For instance, a room with hard surfaces like concrete or glass reflects sound waves, whereas a room with soft surfaces such as carpets or curtains absorbs them. This variance can drastically impact the clarity and quality of the sound we hear. 
% in a well-designed concert hall, the sounds are crisp, clear, and authentic, whereas speaking in an empty live room results in echoes and reverberations, creating a hollow and muffled sound. Additionally, the materials used in constructing the environment can also affect the transmission of sound.

To ensure an authentic and captivating experience, it is imperative to accurately model the acoustics of a room, particularly in virtual reality (VR) and augmented reality (AR) applications. Recent years have seen a surge in significant research~\cite{li2022blip,radford2021learning,li2023blip,huang2023audiogpt} addressing the language-visual modeling problem. For instance, ~\citet{li2022blip} have proposed a unified video-language pre-training framework for learning robust representation, while ~\citet{radford2021learning} have focused on large-scale image-text pairs pre-training via contrastive learning. Visual TTS open-ups numerous practical applications, including dubbing archival films, providing a more immersive and realistic experience in virtual and augmented reality, or adding appropriate sound effects to games.

Despite the benefits of language-visual approaches, training visual TTS models typically requires a large amount of training data, while there are very few resources providing parallel text-visual-audio data due to the heavy workload. Besides, creating a sound experience that matches the visual content remains challenging when developing AR/VR applications, as it is still unclear how various regions of the image contribute to reverberation and how to incorporate the visual modality as auxiliary information in TTS.

In this work, we formulate the task of visual TTS to generate audio with reverberation effects in target scenarios given a text and environmental image, introducing ViT-TTS to address the issues of data scarcity and room acoustic modeling. 
% we leverage ResNet18~\cite{he2016deep} pre-trained on the IMAGENET1K\_V1~\cite{imagenet15russakovsky} dataset to extract image features. Furthermore, with the growing popularity of self-supervised models in the NLP~\cite{devlin2018bert,oord2018representation} and speech processing domains~\cite{hsu2021hubert,schneider2019wav2vec,Chung2021w2vBERTCC}, a significant amount of work has focused on using unlabeled data for pre-training. This work mainly falls into two categories: contrastive loss~\cite{oord2018representation,chung2020improved} and masked prediction loss~\cite{devlin2018bert}. The former involves constructing positive and negative samples to enable the model to differentiate between them, while the latter involves masking the inputs and tasking the model with predicting them. Given that visual TTS is more closely related to masked prediction loss, we have pre-trained the text encoder and diffusion decoder by randomly masking text and mel spectrograms to address low-resource scenarios.
To enhance visual-acoustic matching, we 1) propose the visual-text fusion to integrate visual and textual information, which provides fine-grained language-visual reasoning by attending to regions of the image; 2) leverage transformer architecture to promote the scalability of the diffusion model. Regarding the data shortage challenge, we pre-train the encoder and decoder in a self-supervised manner, showing that large-scale pre-training reduces data requirements for training visual TTS models.
% existing diffusion-based generative models~\cite{popov2021grad,github2021DiffSinger,huang2022prodiff} have demonstrated impressive audio quality for TTS tasks. However, as additional scene information is integrated, a larger model capacity is needed, which is constrained by the convolutional architecture of current diffusion models. Motivated by the recent advances achieved by scalable transformers in image synthesis~\cite{bao2023one,peebles2023scalable}, we propose a diffusion model that possesses the scalability of transformers to replace WaveNet~\cite{oord2016wavenet} and U-Net~\cite{ronneberger2015u} in modeling integrated scene information. Furthermore, we introduce an .To tackle the challenges of data shortage

Experiments results demonstrate that ViT-TTS generates speech samples with accurate reverberation effects in target scenarios, achieving new state-of-the-art results in terms of perceptual quality. In addition, we investigate the scalability of ViT-TTS and its performance under low-resource conditions (1h/2h/5h). The main contributions of this work are summarized as follows:
\begin{itemize}
    \item We propose the first visual Text-to-Speech model ViT-TTS with vision-text fusion, which enables the generation of high-perceived audio that matches the physical environment.
    \item We show that large-scale pre-training alleviates the data scarcity in training visual TTS models.
    \item We introduce the diffusion transformer scalable in terms of parameters and capacity to learn visual scene information. 
    \item Experimental results on subjective and objective evaluation demonstrate the state-of-the-art results in terms of perceptual quality. With low-resource data (1h, 2h, 5h), ViT-TTS achieves comparative results with rich-resource baselines.
\end{itemize}

%% file: sections/related_work.tex
\vspace{-2mm}
\section{Related Work}

\subsection{Text-To-Speech}
Text-to-Speech(TTS) tasks are divided into two categories: (1) generating a mel-spectrogram from text or phoneme sequence first~\cite{wang2017tacotron,ren2019fastspeech}, and then converting the generated spectrum into a waveform via vocoder~\cite{kong2020hifi,lee2022bigvgan,huang2022fastdiff}; (2) generating audio directly from text~\cite{donahue2020end,kim2021conditional}. The earlier TTS~\cite{li2019neural,wang2017tacotron} models adopt an autoregressive manner, which suffers from the problem of slow inference speed. As a solution, non-autoregressive models have been proposed to enable fast inference by generating mel-spectrograms in parallel. More recently, Grad-TTS~\cite{popov2021grad}, DiffSpeech~\cite{github2021DiffSinger}, and ProDiff~\cite{huang2022prodiff} have employed diffusion generative models to generate high-quality audio, but they all rely on the convolutional architecture such as WaveNet~\cite{oord2016wavenet} and U-Net~\cite{ronneberger2015u} as the backbone. In contrast, some studies~\cite{peebles2023scalable,bao2023one} in image generation tasks have explored transformers~\cite{vaswani2017attention} as an alternative to convolutional architectures, achieving competitive results with U-Net. In this paper, we present the first transformer-based diffusion model as an alternative of convolutional architecture. By harnessing the scalable properties of transformers, we enhance the model capacity to more effectively capture visual scene information and promote the model performance.

\subsection{Self-supervised Pre-training}
There are two main criteria for optimizing speech pre-training: contrastive loss~\cite{oord2018representation,chung2020improved,baevski2020wav2vec} and masked prediction loss~\cite{devlin2018bert}. Contrastive loss is used to distinguish between positive and negative samples with respect to a reference sample, while masked prediction loss is originally proposed for natural language processing~\cite{devlin2018bert,Lewis2019BARTDS} and later applied to speech processing~\cite{baevski2020wav2vec,hsu2021hubert}. Some recent work~\cite{Chung2021w2vBERTCC} has combined the two approaches, achieving good performance for downstream automatic speech recognition (ASR) tasks. In this work, we leverage the success of self-supervised to enhance both the encoder and decoder to alleviate the data scarcity issue.
% In this paper, we introduce a new task called visual TTS, which is closely related to masked prediction loss. To tackle the issue of data scarcity in this task, we utilize a large volume of unlabeled data to pre-train both the visual-text encoder and denoiser decoder, which shows that large-scale pre-training alleviates the data scarcity issue in visual TTS systems.

\subsection{Acoustic Matching}
The primary objective of acoustic matching is to convert audio from a source environment into audio that resembles the target environment. In the field of blind estimation~\cite{Mack2020SingleChannelBD,xiong2018joint,Murgai2017BlindEO,Mezghani_2018}, acoustic matching is applied to generate a simple room impulse response (RIR) that can be used to synthesize the corresponding target audio using two critical acoustic metrics - the direct-to-reverberant ratio (DRR)~\cite{zahorik2002direct} and the reverberation time 60 (RT60)~\cite{ratnam2003blind}. 
\begin{figure*}[h]
    \centering
    \vspace{-4mm}
    \includegraphics[scale=0.5]{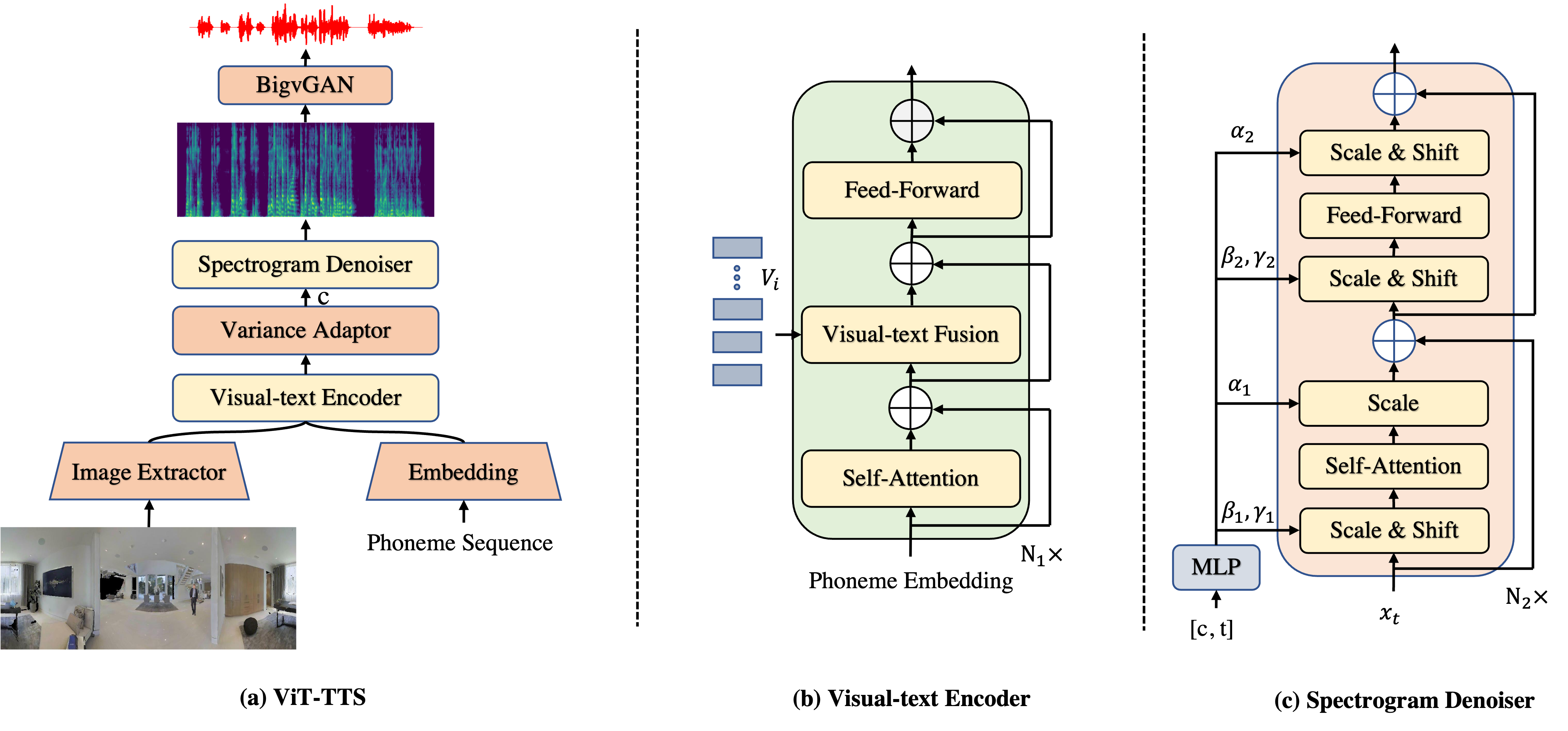}
    \caption{ The overall architecture for ViT-TTS. In subfigure (b), $V_i$ denotes the visual sequence and $N_1$ denotes the layers of Encoder. In subfigure (c), $N_2$ is the number of transformer layers. $\alpha$ and $\beta$  are the dimension-wise scale parameters, while $\gamma$ is the dimension-wise shift parameters. c is the variance adaptor's output and t is the diffusion step.}
    \label{fig:fig2}
    % \vspace{-6mm}
\end{figure*}
% The DRR is used to describe the energy ratio between the direct-to-reverberant sound and the reflected sound, while the RT60 is used to measure the time taken for the sound to decay by 60 dB.
The music production community also implements acoustic matching to modify the reverberation, thus simulating the reverberation of the target space or processing algorithm~\cite{koo2021reverb,sarroff2020blind}. Recently, there is research on visual acoustic matching~\cite{Chen_2022_CVPR}, which involves generating audio recorded in the target environment based on the input source audio clip and an image of the target environment. However, our proposed visual TTS is distinct from those mentioned above as as it aims to generate audio that captures the room acoustics in the target environment based on the written text and the target environment image.

%% file: sections/method.tex
\section{Method}

\subsection{Overview}
The overall architecture has been presented as Figure ~\ref{fig:fig2}. To alleviate the issue of data scarcity, we leverage unlabeled data to pre-train the visual-text encoder and denoiser decoder with scalable transformers in a self-supervised manner. To capture the visual scene information, we employ the visual-text fusion module to reason about how different image patches contribute to texts. BigvGAN~\citep{lee2022bigvgan} converts the mel-spectrograms into audio that matches the target scene as a neural vocoder.

\subsection{Enhanced visual-text Encoder}
\paragraph{\textbf{Self-supervised Pre-training}}
% \textbf{Masked Encoder Pre-training}
The advent of the masked language model \cite{devlin2018bert,clark2020electra} has marked a significant milestone in the field of natural language processing.
% By replacing the conventional left-to-right or right-to-left language models, it has demonstrated remarkable results. 
To alleviate the data scarcity issue~\citep{huang2022transpeech,liu2023wav2sql,huang2023avtranspeech} and learn robust contextual encoder, we are encouraged to adopt the masking strategy like BERT in the pre-training stage. Specifically, we randomly mask the $15\%$ of each phoneme sequence and predict those masked tokens rather than reconstructing the entire input. The masked phoneme sequence is then input into the text encoder to obtain hidden states. The final hidden states are fed into a linear projection layer over the vocabulary to obtain the predicted tokens. Finally, we calculate the cross entropy loss between the predicted tokens and target tokens.

The masked token during the pre-training phase will not be used in the fine-tuning phase. To mitigate this mismatch between the pre-training and fine-tuning, we randomly choose the phonemes to be masked: 1) 80\% probability to add masks; 2) 10\% probability to keep phoneme unchanged, and 3) 10\% probability to replace with a random token in the dictionary.

\paragraph{\textbf{Visual-Text Fusion}}
In the fine-tuning stage, we integrate the visual modal and module into the encoder to integrate visual and textual information. Before feeding into the visual-text encoder, we first extract image features of panoramic images through ResNet18~\citep{oord2018representation} and obtain phoneme embedding. Both the image features and phoneme embedding are fed into one of the variants of the transformer to get the hidden sequences. Specifically, we first pass the phoneme through relative self-attention, which is defined as follows:
\begin{equation}
  \small
  \label{eq:rela-attn}
  \alpha(i,j) = Softmax(\frac{(Q_iW^Q)(K_jW^K + R_{ij})^T}{\sqrt{d_k}})
\end{equation}
where n is the length of phoneme embedding, $R_{ij}$ are the relative position embedding of key and value, $d_k$ is the dimension of key, and Q, K, V are all the phoneme embedding. We use relative self-attention to model how much phoneme $p_i$ attends to phoneme $p_j$. After that, we choose to use cross-attention instead of a simplistic concatenation approach as we can reason about how different image patches contribute to the text after feature extraction.
% This is particularly important because different visual attributes in panoramic images carry varying degrees of significance and contribute differently to the final generated audio. As such, it is crucial to leverage cross-attention to calculate the attention weights of each phoneme on different patches. By doing this, we can gain important insights into how different aspects of visual information contribute to the textual output.
The equation is defined as follows:
\begin{equation}
  \small
  \label{eq:cross}
  \delta(V,P) = Softmax(\frac{PV^T}{\sqrt{d_v}})V
\end{equation}
where P is the phoneme embedding, V is the visual features, and $d_v$ is the dimension of vision features. Finally, the feed-forward layer is applied to output the hidden sequence.

\subsection{Enhanced Diffusion Transformer}
\paragraph{\textbf{Scalable Transformer}}
As a rapidly growing category of generative models, DDPMs have demonstrated their exceptional ability to deliver top-notch results in both image~\citep{zhang2023adding,ho2022classifier} and audio synthesis~\citep{huang2022prodiff,huang2023make,lam2021bilateral}.
However, the most dominant diffusion TTS models adopt a convolutional architecture like WaveNet or U-Net as the de-factor choice of backbone. This architectural choice limits the model scalability to effectively incorporate panoramic visual images.
% This prevents the model's incorporation of visual information as they lack scalability.
Recent research~\citep{peebles2023scalable,bao2023one} in the image synthesis field has revealed that the inductive bias of convolutional structures is not a critical determinant of DDPMs' performance. Instead, transformers have emerged as a viable alternative.

For this reason, we propose a diffusion transformer that leverages the scalability of transformers to expand model capacity and incorporate room acoustic information.
% However, there are several challenges when facing visual TTS tasks: (1)The most dominant diffusion TTS models adopt a convolutional architecture like WaveNet or U-Net as the de-factor choice of backbone, where it lacks the scalable ability to model additional visual information. This prevents the model's incorporation of visual information as they lack scalability. (2)Training visual TTS models typically requires a large amount of parallel training data, while resources providing parallel multimodal data could be limited due to the heavy load.
% In this work, we propose two key techniques to complement the above issues: (1)Utilizing the scalable transformer as the backbone to better model scene features. Recent research in the image synthesis field has revealed that the inductive bias of convolutional structures is not a critical determinant of DDPMs' performance. Instead, transformers have emerged as a viable alternative. 
% In comparison to diffusion models~\cite{popov2021grad,github2021DiffSinger,huang2022prodiff} that use WaveNet or U-Net as a backbone, we propose a transformer-based DDPM that leverages the scalability of transformers to expand model capacity and incorporate room acoustic information.
Moreover, we leverage the adaptive normalization layers in GANs and initialize the full transformer block as the identity function to enhance the transformer architecture. 
% Through the preliminary comparisons of model size in Section~\ref{4.2}, we could conclude that the transformer diffusion we use has advantages in modeling room acoustic information compared with WaveNet.
% (2) Alleviating the issue of data scarcity via self-supervised pre-training, we adopt the masked prediction to pre-train the unconditional decoder following the common practice of Wav2vec 2.0 and show that large-scale pre-training pre-training benefits. Furthermore, we introduce Controllable transformer diffusion to achieve fast learning of input conditions, which further reduces the requirements of large-amount multi-modal data and boosts the performance of visual TTS systems in low-resource scenarios and control diffusion model to learn specific conditions during the fine-tuning stage.
\paragraph{\textbf{Unconditional Pre-training}}
In this part, we investigate self-supervised learning from orders of magnitude mel-spectrograms data to alleviate data scarcity. Specifically, assuming the target mel-spectrogram is $\vx_0$, we first random select 0.065\% of $\vx_0$ as starting indices and apply a mask that spans 10 steps following the Wav2vec2.0~\cite{baevski2020wav2vec}. Then, we obtain $\vx_t$ through a diffusion process, which is defined by a fixed Markov chain from data $\vx_0$ to the latent variable $\vx_t$.
\begin{equation}
    \small
    \label{diffusion}
q(\vx_{1},\cdots,\vx_T|\vx_0) = \prod_{t=1}^T q(\vx_t|\vx_{t-1}),
\quad\ \ 
\end{equation}
At each diffusion step $t\in[1,T]$, a tiny Gaussian noise is added to $\vx_{t-1}$ to obtain $\vx_t$, according to a small positive constant $\beta_t$:
\begin{equation}
  \small
q(\vx_t|\vx_{t-1}) := \gN(\vx_t;\sqrt{1-\beta_t}\vx_{t-1},\beta_t \rmI)
\end{equation}
$x_t$ obtained from the diffusion process is passed through the transformer to predict Gaussian noise $\beps_\theta$. Loss is defined as mean squared error in the $\beps$ space, and efficient training is optimizing a random term of $t$ with stochastic gradient descent:
\begin{equation}
    \small
    \label{eq: loss1}
    \gL_{\theta}^{\text{Grad}} = \left\lVert \beps_\theta\left(\alpha_t\vx_{0}+\sqrt{1-\alpha_t^2}\beps\right)-\beps\right\rVert_2^2, \beps\sim\gN(\vzero, \rmI)
\end{equation}

To this end, ViT-TTS takes advantage of the reconstruction loss to predict the self-supervised representations which largely alleviates the challenges of data scarcity. Detailed formulation of DDPM has been attached in Appendix~\ref{Posterior}.

\paragraph{\textbf{Controllable Fine-tuning}}
During the fine-tuning stage, we will face the following challenges: (1) there is a data scarcity issue with the available panoramic images and target environmental audio for training; (2) a fast training method is equally crucial for optimizing the diffusion model, as it can save a significant amount of time and storage space. To address these challenges, we draw inspiration from ~\citet{zhang2023adding} and implement a swift fine-tuning technique. Specifically, we create two copies of the pre-trained diffusion model weights, namely a "trainable copy" and a "locked copy," to learn the input conditions.  We fix all parameters of the pre-trained transformer, designated as $\Theta$, and duplicate them into a trainable parameter $\Theta_t$. We train these trainable parameters and connect them with the "locked copy" via zero convolution layers.  These convolution layers are unique as they have a kernel size of one by one and weights and biases set to zero, progressively growing from zeros to optimized parameters in a learned fashion. 
% \begin{figure}
%     \centering
%     \vspace{-4mm}
%     \includegraphics[scale=0.4]{figs/control.png}
%     \vspace{-2mm}
%     \caption{Controllable fine-tuning illustration. c is the encoder output.}
%     \vspace{-4mm}
%     \label{fig:control}
% \end{figure}

\subsection{Architecture}
As illustrated in Figure~\ref{fig:fig2}, our model comprises a visual-text encoder, variance adaptor, and spectrogram denoiser. The visual-text encoder converts phoneme embeddings and visual features into hidden sequences, while the variance adaptor predicts the duration of each hidden sequence to regulate the length of the hidden sequences to match that of speech frames. Furthermore, different variances like pitch and speaker embedding are incorporated with hidden sequences following FastSpeech 2~\citet{ren2022fastspeech}. Finally, the spectrogram denoiser iteratively refines the length-regulated hidden states into mel-spectrograms. We put more details in Appendix~\ref{appendix:arch}.

\textbf{Visual-Text Encoder} The visual-text encoder consists of relative position transformer blocks based on the transformer architecture. Specifically, it convolves a pre-net for phoneme embedding, a visual feature extractor for image, and a transformer encoder which includes multi-head self-attention, multi-head cross-attention, and feed-forward layer.

\textbf{Variance Adaptor} In variance adaptor, the duration and pitch predictors share a similar model structure consisting of a 2-layer 1D-convolutional network with ReLU activation, each followed by the layer normalization and the dropout layer, and an extra linear layer to project the hidden states into the output sequence.

\textbf{Spectrogram Denoiser} Spectrogram denoiser takes in $\vx_t$ as input to predict $\beps$ added in diffusion process conditioned on the step embedding $E_t$ and encoder output.
% Previously, WaveNet structure was commonly used as a spectrogram denoiser, mainly composed of convolutional blocks. In this work, we first use a transformer structure as our denoiser, which is based on a feed-forward transformer structure.
We adopt a variant of the transformer as our backbone and make some improvements upon the standard transformer motivated by~\citet{peebles2023scalable}, mainly includes:(1) we explore replacing standard layer norm layers in transformer blocks with adaptive layer norm (adaLN) to regress scale and shift parameters from the sum of the embedding vector of t and hidden sequence. (2) Inspired by ResNets~\cite{oord2018representation}, we initialize the transformer block as the identity function and initialize the MLP to output the zero-vector.

\subsection{Pre-training, Fine-tuning, and Inference Procedures}
\paragraph{\textbf{Pre-training}} The pre-training has two stages: 1) encoder stage: pre-train the visual-text encoder vias masked LM loss $\gL_{CE}$ (ie. cross-entropy loss) to predict the masked tokens. 2) decoder stage: the masked $\vx_0$ is puted into denoiser to predict Gaussian noise $\beps_\theta$. Then, the Mean Square Error(MSE) loss is applied to the predicted Gaussian noise and target Gaussian noise.

\paragraph{\textbf{Fine-tuning}} We begin by loading model weights from the pre-trained visual-text encoder and unconditional diffusion decoder, after which we finetune both of them until the model converges. The final loss term consists of the following parts: (1) sample reconstruction loss $\gL_{\theta}$: MSE between the predicted Gaussian noise and target Gaussian noise. (2) variance reconstruction loss $\gL_{dur}, \gL_{p}$: MSE between the predicted and the target phoneme-level duration, pitch.

\paragraph{\textbf{Inference}} During inference, DDPM iteratively runs the reverse process to obtain the data sample $\vx_0$, and then we use a pre-trained BigvGAN-16khz-80band as the vocoder to transform the generated mel-spectrograms into waveforms.

%% file: sections/experiment.tex
\section{Experiment}
\begin{table*}[ht]
  \centering
  \vspace{-3mm}
  % \small
  \begin{tabular}{l|ccc|ccc|c}
      \toprule
 \bfseries \multirow{2}{*}{Method} & \multicolumn{3}{c|}{\bfseries Test-Seen} & \multicolumn{3}{c}{\bfseries Test-Unseen} & \bfseries \multirow{2}{*}{Params} \\
 & \bfseries MOS($\uparrow$) & \bfseries RTE ($\downarrow$) & \bfseries MCD ($\downarrow$) & \bfseries MOS($\uparrow$) & \bfseries RTE ($\downarrow$) & \bfseries MCD ($\downarrow$) \\
  \midrule
  GT                  &  4.34$\pm$0.07       &  /      & /  & 4.24$\pm$0.07   & /  & / & / \\
  GT (voc.)           &  4.18$\pm$0.05       &  0.006      & 1.46  & 4.19$\pm$0.07  &  0.008  & 1.50  & / \\
  \midrule
  WaveNet             &  3.85$\pm$0.09       &  0.091      & 4.61  & 3.78$\pm$0.12  &  0.110  & 4.69 & 42.3M \\
  Transformer-S       &  3.92$\pm$0.07       &  0.068      & 4.57  & 3.80$\pm$0.06  &  0.077  & 4.68 & 32.38M \\
  Transformer-B       &  3.98$\pm$0.06       &  0.061      & 4.53  & 3.90$\pm$0.07  &  0.066  & 4.62 & 41.36M \\
  Transformer-L       &  4.02$\pm$0.08       & 0.056      & 4.37  & 3.95$\pm$0.07  &  0.061  & 4.50 & 56.96M \\
  Transformer-XL      &  \bfseries 4.05$\pm$0.07       &  \bfseries 0.047      & \bfseries 4.35  & \bfseries4.00$\pm$0.05  &  \bfseries 0.053  & \bfseries 4.39 & 115.12M \\
  
  \bottomrule
  \end{tabular}
  % \vspace{1mm}
  \caption{Comparison between the diffusion WaveNet and diffusion transformers sweeping over model config(S, B, L, XL). All models remove the pre-training stage and other conditions not related to backbone in training and inference remain the same.}
  \vspace{-2mm}
  \label{table:tab1}
  \end{table*}
\subsection{Experimental Setup}
\paragraph{\textbf{Dataset}}
We use the SoundSpaces-Speech dataset~\cite{chen2023learning}, which is constructed on the SoundSpaces platform based on real-world 3D scans to obtain environmental audio.
The dataset includes 28,853/1,441/1,489 samples for training/validation/testing, each consisting of clean text, reverberant audio, and panoramic camera angle images.
Following ~\cite{Chen_2022_CVPR}, we remove out-of-view samples and divide the test set into test-unseen and test-seen, where the unseen set injects room acoustics depicted in novel images while the seen set only contains the scenes we have seen in the training stage. We convert the text sequence into the phoneme sequence with an open-source grapheme-to-phoneme conversion tool~\cite{sun2019token}~\footnote{\url{https://github.com/Kyubyong/g2p}}.

Following the common pratice~\cite{ren2019fastspeech,github2021DiffSinger}, we conduct preprocessing on the speech and text data: 1) extract the spectrogram with the FFT size of 1024, hop size of 256, and window size of 1024 samples; 2) convert it to a mel-spectrogram with 80 frequency bins; and 3) extract F0 (fundamental frequency) from the raw waveform using Parselmouth tool~\footnote{\url{https://github.com/YannickJadoul/Parselmouth}}.
% \paragraph{\textbf{Evaluation Metrics}}
% We measure the sample quality of the generated waveform using both objective metrics and subjective indicators. The objective metrics we collected are designed to measure varied aspects of waveform quality between the ground-truth audio and the generated sample. Following the common practice of ~\cite{huang2022prodiff,github2021DiffSinger,popov2021grad}, we randomly select a part of the test set for objective evaluation, here is 50. We provide the following metrics: (1) \textbf{RT60 Error(RTE)}-the correctness of the room acoustics between the predicted waveform and target waveform's RT60 values. RT60 indicates the reverberation time in seconds for the audio signal to decay by 60 dB, a standard metric to characterize room acoustics. We estimate the RT60 directly from magnitude spectrograms of the output audio, using a model trained with disjoint SoundSpaces data. See appendix~\ref{appendix:rt60} for more details. (2) \textbf{Mel Cepstral Distortion(MCD)}-measures the spectral distance between the synthesized and reference mel-spectrum features. The utilization of RTE is solely intended for evaluating the room acoustic performance of the generated audio, and as an additional measure, we have incorporated the MCD metric to assess the quality of the mel-spectrogram.

% For subjective metrics, we use crowd-sourced human evaluation via Amazon Mechanical Turk, where raters are asked to rate \textbf{Mean Opinion Score(MOS)} on a 1-5 Likert scale. More information on MOS evaluation has been attached in Appendix~\ref{appendix:mos}.

\textbf{Model Configurations}
% We extract the mel-spectrogram from the raw waveform and set the hop size and frame size to 256 and 1024 in respect of the sample rate 16kHz. 
The size of the phoneme vocabulary is 73. The dimension of phoneme embeddings and the hidden size of the visual-text transformer block are both 256. We use the pre-trained ResNet18 as an image feature extractor. As for the pitch encoder, the size of the lookup table and encoded pitch embedding are set to 300 and 256. In the denoiser, the number of transformer-B layers is 5 with the hidden size 384 and head 12.
We initialize each transformer block as the identity function and set T to 100 and $\beta$ to constants increasing linearly from $\beta_1$ = $10^{-4}$ to $\beta_T$ = 0.06.
We have attached more detailed information on the model configuration in Appendix~\ref{appendix:arch}

\textbf{Pre-training, Fine-tuning, and Inference} During the pre-training stage,
% we set the mask probability of our encoder to 15\% and the mask span of our decoder to 10 with q=0.065.
we pre-train the encoder for 120k steps and the decoder for 160k until convergence. The diffusion probabilistic models have been trained using 1 NVIDIA A100 GPU with a batch size of 48 sentences. In the inference stage, we uniformly use a pre-trained BigvGAN-16khz-80band~\citep{lee2022bigvgan} as a vocoder to transform the generated mel-spectrograms into waveforms.

% \textbf{Baselines} We compare our model with the following baselines:
% \begin{itemize}
%   \item \textbf{GT} the ground-truth audio;
%   \item \textbf{GT(voc.)} we first convert the ground-truth audio into mel-spectrograms and then convert them back to audio using BigvGAN;
%   \item \textbf{DiffSpeech~\cite{github2021DiffSinger}} a denoising diffusion probabilistic model based on WaveNet.
%   \item \textbf{ProDiff~\cite{huang2022prodiff}} a recent generator-based diffusion model proposed to reduce the sampling time.
%   \item \textbf{Visual DiffSpeech} Incorporate visual-text fusion module into DiffSpeech
%   \item \textbf{Cascaded} the system composed of DiffSpeech and Visual Acoustic Matching(VAM). The generated audio of DiffSpeech is then input into the VAM as the source audio.
% \end{itemize}

\begin{table*}[ht]
  \centering

    \begin{tabular}{l|ccc|ccc|c}
    \toprule
    \multirow{2}{*}{\bfseries Method} & \multicolumn{3}{c|}{\bfseries Test-Seen} & \multicolumn{3}{c}{\bfseries Test-Unseen} & \bfseries \multirow{2}{*}{Params} \\
    & \bfseries MOS ($\uparrow$) & \bfseries RTE ($\downarrow$) & \bfseries MCD ($\downarrow$) & \bfseries MOS ($\uparrow$) & \bfseries RTE ($\downarrow$) & \bfseries MCD ($\downarrow$) \\
    \midrule
    GT                                              & 4.34$\pm$0.07   & /   &  / &  4.24$\pm$0.07  &  / & / & / \\
    GT(voc.)                                        & 4.18$\pm$0.05   & 0.006   &  1.46 &  4.19$\pm$0.07  &  0.008& 1.50 & / \\
    \midrule            
    DiffSpeech            & 3.79$\pm$0.08   & 0.104   &  4.65 &  3.67$\pm$0.05  &  0.120 & 4.71 & 29.9M \\
    ProDiff                 & 3.76$\pm$0.13   & 0.121   &  4.67 &  3.65$\pm$0.06  &  0.137 & 4.72 & 29.9M \\
    Visual-DiffSpeech           & 3.85$\pm$0.09   & 0.091   &  4.61 &  3.78$\pm$0.12  &  0.110& 4.69 & 42.3M \\
    Cascaded                                        & 3.61$\pm$0.08   & 0.071   &  5.13 &  3.59$\pm$0.08  &  0.082& 5.25 & 146.5M \\
    \midrule     
    \bfseries ViT-TTS                                         & \bfseries 3.95$\pm$0.06   & \bfseries 0.066   &  \bfseries 4.52 &  \bfseries 3.86$\pm$0.05  &  \bfseries 0.076& \bfseries 4.59 & 41.3M \\
    \bottomrule
    \end{tabular}
    % \vspace{1mm}
    \caption{Comparison with baselines on the SoundSpaces-Speech for Seen and Unseen scenarios. The diffusion step of all diffusion models is set to 100. We use the pre-trained model provided by VAM for the evaluation of cascaded.}
    \vspace{-4mm}
    \label{table:overall}
    \end{table*}  

\subsection{Scalable Diffusion Transformer}
\label{4.2}
We compare and examine diffusion transformer sweeping over model config(S, B, L, XL), and conduct evaluations in terms of audio quality and parameters. Appendix~\ref{appendix:config} gives the details of the model configs. The results have been shown in Table ~\ref{table:tab1}. We have some observations from the results: (1) Increasing the depth and number of layers in the transformer can significantly enhance the performance of the diffusion model, resulting in an improvement in both objective metrics and subjective metrics, which demonstrates that expanding the model size enables finer-grained room acoustic modeling.
% This is evidenced by the decrease in the RTE from 0.077 to 0.053 on the test-unseen set and from 0.068 to 0.047 on the test-seen set, which demonstrates that expanding the model capability could better model the room acoustic information and boost the performance of diffusion models.
(2) Our proposed diffusion transformer outperforms WaveNet backbone under similar parameters across both test-unseen and test-seen sets, significantly in the rt60 metric.
% Our proposed diffusion transformers achieve audio sampling quality slightly higher than diffusion WaveNet and surpasses it significantly in the rt60 metric, proving the superior modeling capability of transformer in room acoustics.
We attribute this to the fact that instead of directly concatenating the condition input like WaveNet, we replace standard layer norm layers in transformer blocks with adaptive layer norm to regress dimension-wise scale and shift parameters from the sum of the embedding vectors of diffusion step and encoder output, which can better incorporate the conditional information, as proven in GANs~\cite{brock2018large,karras2019style}.

\vspace{-2mm}
\subsection{Model Performances}
In this study, we conduct a comprehensive comparison of the generated audio quality with other systems, including 1) GT, the ground-truth audio; 2) GT(voc.), where we first convert the groud-truth audio into mel-spectrograms and then convert them to audio using BigvGAN; 3) DiffSpeech~\citep{github2021DiffSinger}, one of the most popular DDPM based on WaveNet; 4)ProDiff~\citep{huang2022prodiff}, a recent generator-based diffusion model proposed to reduce the sampling time; 5)Visual-DiffSpeech, incorporate visual-text fusion module into DiffSpeech; 6) Cascaded, the system composed of DiffSpeech and Visual Acoustic Matching(VAM)~\citep{Chen_2022_CVPR}. The results, compiled and presented in Table~\ref{table:overall}, provide valuable insights into the effectiveness of our approach:

(1) As expected, the results in the test-unseen set do poorer than the test-seen part because there are invisible scenarios among the test-unseen set. However, our proposed model has achieved the best performance compared to baseline systems in both sets, indicating that our model generates the best-perceived audio that matches the target environment from written text. 
(2) Our model surpassed TTS diffusion models(i.e.DiffSpeech and ProDiff) across all metric scores, especially in terms of RTE values. This suggests that conventional diffusion models in TTS do poorly in modeling room acoustic information, as they mainly focus on audio content, pitch, energy, etc. Our proposed visual-text fusion module addresses this challenge by injecting visual properties into the model, resulting in a more accurate prediction of the correct acoustics from images and high-perceived audio synthesis. 
(3) The results of comparison with Visual-DiffSpeech highlight the advantages of our choice of transformer and self-supervised pre-training. Although Visual-DiffSpeech adds the visual-text module, the choice of WaveNet and the lack of a self-supervised pre-training strategy make it perform worse in predicting the correct acoustics from images and synthesizing high-perceived audio. 
(4) The cascaded system composed of DiffSpeech and Visual Acoustic Matching model visual properties is better than other baselines. However, compared to our proposed model, it performed worse in both test-unseen and test-seen environments. This suggests that our direct visual text-to-speech system eliminates the influence of error propagation caused by the cascaded manner, resulting in high-perceived audio.
In conclusion, our comprehensive evaluation results demonstrate the effectiveness of our proposed model in generating high-quality audio that matches the target environment.

\begin{figure*}[ht]
    \centering
    \small
    \vspace{-3mm}
    \includegraphics[width=.995\textwidth]{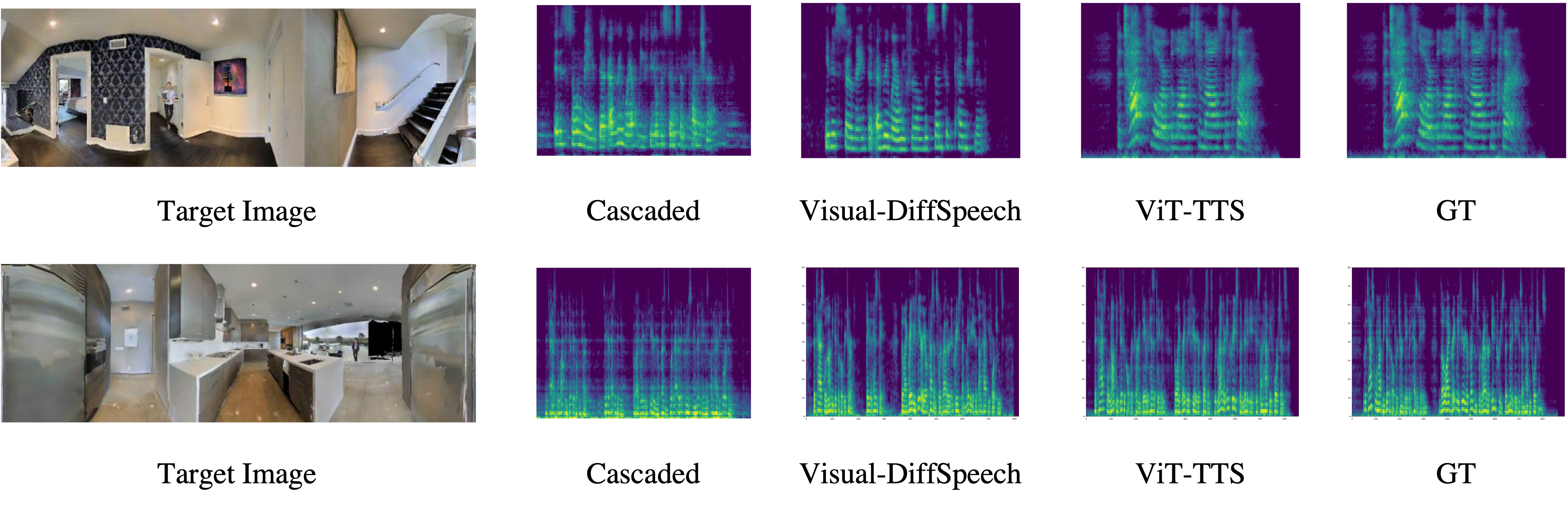}
    \vspace{-2mm}
    \caption{Visualizations of the ground truth and generated mel-spectrograms by different Visual TTS models. The text corresponding to the first line in test-seen is "it is so made that everywhere we feel the sense of punishment" while the second line in test-unseen is "the task will not be difficult returned david hesitating though i greatly fear your presence would rather increase than mitigate his unhappy fortunes
".}
    \vspace{-4mm}
    \label{fig:vis_mel}
    \end{figure*}

\vspace{-2mm}
\subsection{Low Resource Evaluation}
Training visual text-to-speech models typically requires a large amount of parallel target environment image and audio training data, while there may be very few resources due to the heavy workload. In this section, we prepare low-resource audio-visual data (1h/2h/5h) and leverage large-scale text-only and audio-only data to boost the performance of the visual TTS system, to investigate the effectiveness of our self-supervised learning methods. The results are compiled and presented in Table~\ref{table:low}, and we have the following observations: 
1)As training data is reduced in the low-resource scenario, a distinct degradation in generated audio quality could be witnessed in both test sets (test-seen and test-unseen). 2) Leveraging orders of magnitude text-only and audio-only data with self-supervised learning, the ViT-TTS achieve RTE scores of 0.082 and 0.068 respectively in test-unseen and test-seen, showing a significant promotion regardless of the unseen scene. In this way, the dependence on a large number of parallel audio-visual data can be reduced for constructing visual text-to-speech systems.

\begin{table}[h]
  \centering
  \vspace{-2mm}
  \begin{tabular}{lccc}
  \toprule
  \bfseries Method  &\bfseries MOS ($\uparrow$)&\bfseries RTE ($\downarrow$) &\bfseries MCD ($\downarrow$) \\
  \midrule
   \multicolumn{4}{l}{\bfseries Finetune with 1 hour data}    \\
  \midrule
  Test-Seen                         &3.72$\pm$0.05  & 0.092  & 5.04 \\
  Test-Unseen                          &3.67$\pm$0.06  & 0.101  & 5.11 \\
  \midrule
  \multicolumn{4}{l}{\bfseries Finetune with 2 hours data}    \\
  \midrule
  Test-Seen                         &3.75$\pm$0.06  & 0.089  & 4.85 \\
  Test-Unseen                          &3.70$\pm$0.07  & 0.097  & 4.89 \\
  \midrule
   \multicolumn{4}{l}{\bfseries Finetune with 5 hours data}    \\
  \midrule
  Test-Seen                         &3.83$\pm$0.05  & 0.068  & 4.65 \\
  Test-Unseen                          &3.73$\pm$0.09  & 0.082  & 4.72 \\
  \bottomrule
  \end{tabular}
  % \vspace{2mm}
  \caption{Low resource evaluation results.}
  \vspace{-3mm}
  \label{table:low}
  \end{table}

\vspace{-2mm}
\subsection{Case Study}
We provide two examples of generation sampled from a large empty room with significant reverberation in the Test-Seen environment depicted in Figure~\ref{fig:vis_mel}, and have the following observations: 
1) Mel-spectrograms produced by ViT-TTS are noticeably more similar to the target counterpart.
2) Moreover in challenging scenarios with invisible scene images, cascaded systems suffer severely from the issue of noisy and reverb details missing, which is largely alleviated in ViT-TTS.
% 1) ViT-TTS ten
% and our findings show that our proposed model, ViT-TTS, outperforms the benchmark models, Cascaded and Visual-DiffSpeech, in two aspects. Firstly, the complemental visual information incorporated in our model leads to mel-spectrograms that are more similar to the target counterpart. Secondly, as demonstrated by the more uniform and continuous distribution of reflected energy and harmonics in the spectrum generated by ViT-TTS, our model not only achieves better audio quality but also exhibits a superior reverberation effect. Comparatively, Visual-DiffSpeech's mel-spectrogram is overly smoothed, missing fine reverb details, while Cascaded's spectrogram is noisier and lacks content, causing perceptual distortion.

\subsection{Ablation Studies}
We conduct ablation studies to demonstrate the effectiveness of several key techniques on the Test-Unseen set in our model, including the encoder pre-training(EP), decoder pre-training(DP), visual input, random image, and concat function. The results of both subjective and objective evaluations have been presented in Table~\ref{table:ablation}, and we have the following observations: 1) Removing the self-supervised encoder and decoder pre-training strategy results in a decline in all indicators, which demonstrates the effectiveness and efficiency of the proposed pre-training strategy in reducing data variance and promoting model convergence. 2) Without the input of RGB-D image and removing all of the modules related to the image causes a distinct degradation in RTE values, which demonstrates that our model successfully learns acoustics from the visual scene. 3) The replacement of cross-attention with the concat fusion function results in a decrease in performance across all metrics, highlighting the effectiveness of our visual-text fusion module.
\begin{table}[h]
  \centering
  \vspace{-2mm}
  \begin{tabular}{lccc}
  \toprule
  \bfseries Method  &\bfseries MOS ($\uparrow$)&\bfseries RTE ($\downarrow$) &\bfseries MCD ($\downarrow$) \\
  \midrule
  GT(voc.)                        & 4.18$\pm$0.07  & 0.008  & 1.50 \\
  \midrule
  ViT-TTS                         & \bfseries 3.86$\pm$0.05  & \bfseries 0.076  & \bfseries 4.59 \\
  w/o EP                          & 3.82$\pm$0.07  & 0.078  & 4.63 \\
  w/o DP                          & 3.83$\pm$0.06  & 0.081  & 4.65 \\
  w/o Visual                      & 3.78$\pm$0.07  & 0.102  & 4.68 \\
  w/ RI                 & 3.73$\pm$0.08  & 0.103  & 4.75 \\
  w/ Concat          & 3.80 $\pm$0.06 & 0.089 & 4.63 \\
  % \midrule
  \bottomrule
  \end{tabular}
  % \vspace{2mm}
  \caption{Ablation study results. EP, DP, and RI are encoder pre-training, decoder pre-training, and random images respectively. }
  \vspace{-4mm}
  \label{table:ablation}
  \end{table}
  
Furthermore, we conducted a more detailed exploration of our model's processing and reasoning about different patches in the RGB-D images. To achieve this, we deliberately substituted the target image with random images, allowing us to determine whether the model can derive meaningful representations from visual inputs. Our findings show that after replacing the target image with a random image, the performance of our model significantly degraded, indicating that our model could model the room acoustic information of visual input.

% In summary, our study demonstrates the effectiveness of the proposed pre-training strategy and the choice of the scalable transformer.

%% file: sections/conclusion.tex
\vspace{-2mm}
\section{Conclusion}
In this paper, we proposed ViT-TTS, the first visual text-to-speech synthesis model that aimed to convert written text and target environmental images into audio that matches the target environment. 
% ViT-TTS complemented the phoneme sequence with the visual information to generate high-perceived audio, opening up new avenues for practical applications of AR and VR.
% , as it allows for a more immersive and realistic audio experience.
To mitigate the data scarcity for training visual TTS tasks and model visual acoustic information, we 1) introduced a self-supervised learning framework to enhance both the visual-text encoder and denoiser decoder; 2) leveraged the diffusion transformer scalable in terms of parameters and capacity to improve performance.

% Experimental results demonstrated that ViT-TTS achieved state-of-the-art, outperforming cascaded systems regardless of the visibility of the scene.  Additionally, The preliminary study on scaling model size found that the diffusion transformer produced better results than the diffusion WaveNet under the same parameters. With low-resource data (1h, 2h, 5h), ViT-TTS yielded a significant improvement and achieved comparative results with baselines.
Experimental results demonstrated that ViT-TTS achieved new state-of-the-art results and performed comparably to rich-resource baselines even with limited data.
% outperforming cascaded systems and other baselines regardless of the visibility of the scene. With low-resource data (1h, 2h, 5h), ViT-TTS achieves comparative results with rich-resource baselines.
To this end, ViT-TTS provided a solid foundation for future visual text-to-speech studies, and we envision that our approach will have far-reaching impacts on the fields of AR and VR.

% \clearpage
\section{Limitation and Potential Risks}
As indicated in the experimental setup, we utilized ResNet-18 as our image feature extractor. While it is a classic extractor, there may be newer extractors that perform better. In future work, we will explore the use of superior extractors to enhance the quality of generated audio.

Moreover, our pre-trained encoder and decoder are based on the SoundSpace-Speech dataset, which, as described in the dataset section, is not sufficiently large. To address this limitation in future work, we will pre-train on a large-scale dataset to achieve better performance in low-resource scenarios.

ViT-TTS lowers the requirements for visual text-to-speech generation, which may cause fraud and scams by impersonating someone else's voice. Furthermore, there is the potential for leading to the spread of false information and rumors.

\section*{Acknowledgements}

This work was supported in part by the National Natural Science Foundation of China under Grant No.61836002 and Ant Group Research Fund.

%% file: sections/appendix.tex
\section{TRANSFORMER CONFIGURATION} \label{appendix:config}
The details of transformer denoisers are shown in Table~\ref{tab:config}, while B, M, L, and XL means the base, medium, large, extra large respectively.
\begin{table}[h]
    \centering
    \begin{tabular}{cccc}
    \toprule
    Model & layers & Hidden Size & Heads \\ 
    \midrule
    Transformer-S & 4 & 256 & 8 \\
    Transformer-B & 5 & 384 & 12 \\
    Transformer-L & 6 & 512 & 16 \\
    Transformer-XL & 8 & 768 & 16 \\
    \bottomrule
    \end{tabular}
    \caption{Diffusion Transformer Configs.}
    \vspace{-4mm}
    \label{tab:config}
\end{table}

\section{ARCHITECTURE} \label{appendix:arch}
We list the model hyper-parameters of ViT-TTS in Table~\ref{tab:hyperparameters_ps}.

\begin{table}[h]
\small
\centering
\begin{tabular}{l|c|c}
\toprule
\multicolumn{2}{c|}{Hyperparameter}   & ViT-TTS \\ 
\midrule
\multirow{9}{*}{Visual-Text Encoder} 
&Phoneme Embedding           &256   \\
&Pre-net Layers              &3   \\
&Pre-net Hidden              &256   \\
&Visual Conv2d Kernel        &(7, 7) \\
&Visual Conv2d Stride        &(2, 2) \\
&Encoder Layers              &4   \\
&Encoder Hidden              &256     \\                      
&Encoder Conv1d Kernel       &9   \\    
&Conv1D Filter Size  &1024\\                 
&Attention Heads     &2   \\    
&Dropout             &0.1\\                       
\midrule
\multirow{3}{*}{Variance Predictor}         
&Conv1D Kernel        & 3\\    
&Conv1D Filter Size   & 256  \\    
&Dropout              & 0.5  \\ 
\midrule
\multirow{5}{*}{Denoiser}     
&Diffusion Embedding                &  384  \\   
&Transformer Layers                    &  5 \\       
&Transformer Hidden                  &  384 \\     
&Attention Heads              &  12 \\      
&Position Embedding                  & 384 \\
&Scale/Shift Size                    & 384 \\

\midrule
\multicolumn{2}{c|}{Total Number of Parameters}   & 41.36M  \\
\bottomrule
\end{tabular}
\caption{Hyperparameters of ViT-TTS models.}
\label{tab:hyperparameters_ps}
\end{table}

\section{DIFFUSION POSTERIOR DISTRIBUTION} \label{Posterior}
Firstly we compute the corresponding constants respective to diffusion and reverse process:
\begin{align}
    \alpha_{t}=\prod_{i=1}^{t} \sqrt{1-\beta_{i}} \quad \sigma_t=\sqrt{1-\alpha_t^2}
\end{align}

The Gaussian posterior in diffusion process is defined through the Markov chain, where each iteration adds Gaussian noise.
\begin{align}
    \begin{split}
q(\vx_{1},\cdots,\vx_T|x_0) &= \prod_{t=1}^T q(\vx_t|\vx_{t-1}),\\ q(\vx_t|\vx_{t-1})=& \gN(\vx_t;\sqrt{1-\beta_t}\vx_{t-1},\beta_t \rmI)
\quad\ \ 
\end{split}
\end{align}

We emphasize the property observed by~\cite{ho2020denoising}, the diffusion process can be computed in a closed form:
\begin{equation}
    q(\vx_{t}|\vx_{0})=\gN(\vx_{t} ; \alpha_{t} \vx_{0},\sigma_t \rmI)
\end{equation}

Applying Bayes' rule, we can obtain the forward process posterior when conditioned on $\vx_0$
\begin{align}
    \begin{split}
    q(\vx_{t-1}|\vx_{t}, \vx_{0}) &=\frac{q(\vx_{t}|\vx_{t-1}, \vx_{0}) q(\vx_{t-1}|\vx_{0})}{q(\vx_{t}|\vx_{0})} \\
    & =\gN(\vx_{t-1};\tilde{\boldsymbol{\mu}}_{t}(\vx_{t}, \vx_{0}), \tilde{\beta}_{t} \rmI),
\end{split}
\end{align}
where $\tilde{\boldsymbol{\mu}}_{t}(\vx_{t}, \vx_{0})=\frac{\alpha_{t-1} \beta_{t}}{\sigma_t} \vx_{0}+\frac{\sqrt{1-\beta_{t}}(\sigma_{t-1})}{\sigma_t} \vx_{t},  \quad \tilde{\beta}_{t}=\frac{\sigma_{t-1}}{\sigma_t} \beta_{t}$

\section{DIFFUSION ALGORITHM}
See Algorithm~\ref{alg: training2} and ~\ref{alg: sampling}.
\begin{algorithm}[h]
    \centering
    \caption{Training procedure}\label{alg: training2}
    \begin{algorithmic}[1]
     \STATE \textbf{Input}: The denoiser $\epsilon_\theta$, diffusion step T and variance condition c.
    \REPEAT 
    \STATE Sample $\vx_{0} \sim q_{data}$, $\rvepsilon\sim\gN(\vzero,\mI)$
    \STATE Take gradient descent steps on $\nabla_{\theta}||\epsilon - \epsilon_\theta(\sqrt{\overline{\alpha}_t} \vx_{0} + \sqrt{1-\overline{\alpha}_t}\epsilon, c, t)||$. 
    \UNTIL{convergence}
    \end{algorithmic}
    \end{algorithm}
    % \vspace{-2mm}
\begin{algorithm}[h]
  \centering
  \caption{Sampling}\label{alg: sampling}
  \begin{algorithmic}[1]
    \STATE \textbf{Input}: The denoiser $\epsilon_\theta$, and variance condition $c$.
  \STATE Sample $\vx_{T} \sim \gN(\vzero,\mI)$
  \FOR{$t=T,\cdots,1$}
  \IF{t = 1}  
  \STATE z = 0
  \ELSE 
  \STATE Sample $\vz \sim \gN(\vzero,\mI)$
  \ENDIF
  \STATE Sample $\vx_{t-1} = \frac{1}{\sqrt{\alpha_t}}(\vx_t - \frac{1-\alpha_t}{\sqrt{1-\overline{\alpha}_t}}\epsilon_\theta(\vx_t,c,t)) + \sigma_t \vz$
  \ENDFOR
  % \RETURN $\vx_0$
  \end{algorithmic}
  \end{algorithm}
  \vspace{-2mm}

\section{EVALUATION MATRIX} \label{appendix:evaluation}
\begin{figure*}[h]
    \centering
    \vspace{-4mm}
    \includegraphics[scale=0.28]{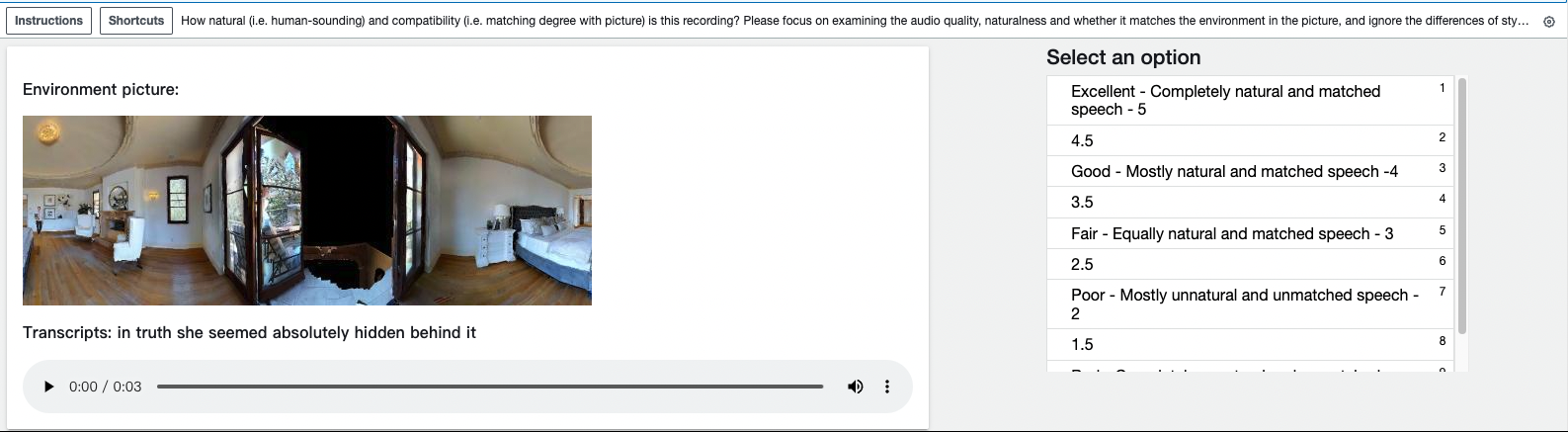}
    \caption{Screenshots of subjective evaluations.}
    \label{fig:screenshot}
    \vspace{-4mm}
\end{figure*}
\subsection{Evaluation Metrics}
We measure the sample quality of the generated waveform using both objective metrics and subjective indicators. The objective metrics we collected are designed to measure varied aspects of waveform quality between the ground-truth audio and the generated sample. Following the common practice of ~\cite{huang2022prodiff,github2021DiffSinger,popov2021grad}, we randomly select a part of the test set for objective evaluation, here is 50. We provide the following metrics: 
(1) \textbf{RT60 Error(RTE)}-the correctness of the room acoustics between the predicted waveform and target waveform's RT60 values. RT60 indicates the reverberation time in seconds for the audio signal to decay by 60 dB, a standard metric to characterize room acoustics. We estimate the RT60 directly from magnitude spectrograms of the output audio, using a model trained with disjoint SoundSpaces data.
% See appendix~\ref{appendix:rt60} for more details.
(2) \textbf{Mel Cepstral Distortion(MCD)}-measures the spectral distance between the synthesized and reference mel-spectrum features. The utilization of RTE is solely intended for evaluating the room acoustic performance of the generated audio, and as an additional measure, we have incorporated the MCD metric to assess the quality of the mel-spectrogram.

For subjective metrics, we use crowd-sourced human evaluation via Amazon Mechanical Turk, where raters are asked to rate \textbf{Mean Opinion Score(MOS)} on a 1-5 Likert scale.
\subsection{RT60 Estimator}\label{appendix:rt60}
Following ~\cite{Chen_2022_CVPR}, we first encode the 2.56s speech clips as spectrograms, process them with a ResNet18~\cite{oord2018representation} and predict the
RT60 of the speech. The ground truth RT60 is calculated
with the Schroeder~\cite{schroeder1965new}. We optimize the MSE loss
between the predicted RT60 and the ground truth RT60.
\subsection{MOS Evaluation}\label{appendix:mos}
To probe audio quality, we conduct the MOS (mean opinion score) tests and explicitly instruct the raters to “focus on
examining the audio quality, naturalness and whether the audio matches with the given image.”. The testers present and rate the samples, and each tester is asked to evaluate the subjective naturalness on a 1-5 Likert scale.

Our subjective evaluation tests are crowd-sourced and conducted via Amazon Mechanical Turk. These ratings are obtained independently for model samples and reference audio, and both are reported. The screenshots of instructions for testers have been shown in Figure~\ref{fig:screenshot}. A small subset of speech samples used in the test is available at \url{https://ViT-TTS.github.io/}

\section{LOW RESOURCE SETTING}
We partition the training set of SoundSpaces-Speech into 1h/2h/5h subsets based on the alphabetical order of speech IDs. Subsequently, we employ these subsets to fine-tune our pre-trained models and assess their performance on identical test sets.

%%
%% End of file `sample-sigconf.tex'.